\documentclass[10pt,letterpaper]{article}
\usepackage[top=0.85in,left=2.75in,footskip=0.75in,marginparwidth=2in]{geometry}

\usepackage[utf8]{inputenc}

\usepackage{cite}

\usepackage{textcomp,marvosym}

\usepackage{amsmath}
\usepackage{textgreek}

\usepackage{nameref,hyperref}

\usepackage[right]{lineno}

\usepackage{microtype}
\DisableLigatures[f]{encoding = *, family = * }

\raggedright
\usepackage{changepage}

\usepackage[aboveskip=1pt,labelfont=bf,labelsep=period,singlelinecheck=off]{caption}

\makeatletter
\renewcommand{\@biblabel}[1]{\quad#1.}
\makeatother

\usepackage{lastpage,fancyhdr,graphicx}
\usepackage{epstopdf}
\fancyheadoffset[L]{2.25in}
\fancyfootoffset[L]{2.25in}

\usepackage{color}

\definecolor{Gray}{gray}{.25}

\usepackage{graphicx}

\usepackage{sidecap}

\usepackage{wrapfig}
\usepackage[pscoord]{eso-pic}
\usepackage[fulladjust]{marginnote}
\reversemarginpar

\begin{document}
\vspace*{0.35in}

\begin{flushleft}
{\Large
\textbf\newline{Sea-level projections representing deeply uncertain ice-sheet contributions}
}
\newline
\\
Alexander M.R. Bakker\textsuperscript{1,2,*},
Tony E. Wong\textsuperscript{1},
Kelsey L. Ruckert\textsuperscript{1},
and Klaus Keller\textsuperscript{1,3,4}
\\
\bigskip
\bf{1} Earth and Environmental Systems Institute, Pennsylvania State University, University Park, PA 16802, USA
\\
\bf{2} Rijkswaterstaat, Ministry of Infrastructure and Environment, Utrecht, Netherlands
\\
\bf{3} Department of Geosciences, Pennsylvania State University, University Park, PA 16802, USA
\\
\bf{4} Department of Engineering and Public Policy, Carnegie Mellon University, Pittsburgh, PA 15289, USA
\\
\bigskip
* Corresponding author: Alexander M.R. Bakker (bakker@psu.edu / alexander.bakker@rws.nl)

\end{flushleft}

\section*{Key points:}
\begin{enumerate}
  \item We present calibrated projections of sea-level rise that represent uncertainties in global temperature changes, oceanic heat uptake, and sea-level rise contributions from thermal expansion, and mass loss from the Antarctic ice sheet, the Greenland ice sheet and small glaciers and ice caps.
  \item These probabilistic sea-level projections explicitly approximate the deeply uncertain contribution of the West Antarctic ice sheet. The projections aim to inform robust decisions, i.e., that are less sensitive to non-trivial or controversial assumptions.
  \item We show how the deep uncertainties surrounding the WAIS contributions can dominate the uncertainties within decades from now.
\end{enumerate}

\section*{Abstract}
Future sea-level rise poses nontrivial risks for many coastal communities~\cite{nicholls2010,hinkel2016}. Managing these risks often relies on consensus projections like those provided by the IPCC~\cite{church2013}. Yet, there is a growing awareness that the surrounding uncertainties may be much larger than typically perceived~\cite{bakker2016b}. Recently published sea-level projections appear widely divergent and highly sensitive to non-trivial model choices~\cite{bakker2016b} and the West Antarctic Ice Sheet (WAIS) may be much less stable than previously believed, enabling a rapid disintegration~\cite{pollard2015,deconto2016}. In response, some agencies have already announced to update their projections accordingly~\cite{KNMI2016web,UNFCCC2016web}. Here, we present a set of probabilistic sea-level projections that approximate deeply uncertain WAIS contributions. The projections aim to inform robust decisions by clarifying the sensitivity to non-trivial or controversial assumptions. We show that the deeply uncertain WAIS contribution can dominate other uncertainties within decades. These deep uncertainties call for the development of robust adaptive strategies~\cite{haasnoot2013}.These decision-making needs, in turn, require mission-oriented basic science, for example about potential signposts and the maximum rate of WAIS induced sea-level changes. 

\paragraph{}
The construction of sea-level projections is often largely motivated by scientific considerations, i.e. in order to better understand the underlying physics~\cite{bakker2015-phd,hinkel2016}. In this process, the translation from input data to model projections and full uncertainty estimates involves a wide range of non-trivial model choices and assumptions than can result in large discrepancies between different uncertainty estimates~\cite{bakker2016b}. For example, many studies consider a high level of model detail indispensable for reliable projections~\cite{church2013} whereas semi-empirical model approaches~\cite{rahmstorf2007,grinsted2010,mengel2016} trade complexity for the ability to calibrate the model. Semi-empirical modeling approaches often rely on strong assumptions about the prior parameter distributions, what mechanisms to include, and how to interpret and represent the data-model discrepancies. Those modeling choices can be nontrivial and the associated uncertainties hard to quantify~\cite{little2013b}. One the other hand, projections based on multi-model ensembles (implicitly) focus on structural uncertainty which requires strong assumptions on which part of the overall uncertainty is covered~\cite{bakker2016b}.
\paragraph{}
Decision makers often prefer "robust" over optimal decisions when faced with "deep" uncertainty~\cite{ellsberg1961,lempert2007,hall2012,herman2015}. Deep uncertainty refers to a situation when experts cannot agree upon or are not willing to provide probabilistic uncertainty ranges~\cite{lempert2007}. In the context of decision making, robustness has many different definitions that usually involve trading some optimality for relative insensitivity to deviations from the model assumptions or relative good performance over a wide range of futures (e.g.~\cite{lempert2007,hall2012,mcinerney2012,herman2015}).
\paragraph{}
Here we present sea-level projections that are intended to support the design of robust strategies to cope with the deep uncertainties surrounding sea-level change, i.e. “solutions capable of withstanding from deviations of the conditions for which they are designed”~\cite{herman2015}. This notion of “robustness” deviates from \textit{scientific} robustness that builds on arguably well understood physics and empirical/robust evidence~\cite{pirtle2010,vellinga2009,vezer2016}, which may lead to overconfident uncertainty ranges~\cite{bakker2016b,hinkel2016} and getting surprised by new insights and data~\cite{bakker2015-phd}.
\paragraph{}
Our sea-level projections are constructed to support robust decision frameworks by i) being explicit about the relevant uncertainties, both shallow and deep; ii) communicating plausible ranges of sea-level rise, including the deep uncertainties surrounding future climate forcings and potential WAIS collapse; and iii) erring on the side of underconfident versus overconfident when possible.

\section*{Model design}
\paragraph{}
We design the projections to be probabilistic where reasonable and explicit about deep uncertainties (e.g. resulting from non-trivial model choices) when needed. Robust decision frameworks often apply plausible rather than probabilistic ranges to represent and communicate uncertainties (e.g.~\cite{ranger2013,herman2015}). In the case of sea-level projections, the bounding of the plausible range usually involves both a probabilistic interpretation of the surrounding uncertainties and estimates of which probabilities are still relevant. For example, a full disintegration of the major ice-sheets is often not taken into account because the probabilities of this occurring are considered too small to be relevant~\cite{church2013,KNMI2014}. What probability is relevant is highly dependent on the decision context and therefore it makes sense to be explicit about the probabilities. Moreover, probabilities are the easiest and most unambiguous way to communicate uncertainties (e.g.~\cite{budescu2012,cooke2015a}).
\paragraph{}
Our projections are designed to highlight the relatively large deep uncertainties, notably those resulting from future climate forcings and those surrounding potential WAIS collapse (even though representations of deep uncertainty often implicitly encompass probabilistic interpretations). The future climate forcing is, to a large extent, controlled by future human decisions.
\paragraph{}
The probability of a WAIS collapse is potentially much larger than previously thought due to the combined effects of Marine Ice Sheet Instability (MISI), ice cliff failure and hydrofracturing~\cite{pollard2015,deconto2016}. The discovery of this new mechanism puts earlier expert elicitations in a different light as it is unclear if those were based on this combined effect. One approach when faced with deeply uncertain model structures and priors is to present a potential WAIS collapse as deeply uncertain by means of a plausible range. We stress that this range is not meant to represent an implicit probabilistic projection of the WAIS contribution to sea-level rise.
\paragraph{}
We merge some small deep uncertainties into the probabilistic part of the projections. According to Herman et al.~\cite{herman2015} "... a larger risk lies in sampling too narrow a range (thus ignoring potentially important vulnerabilities) rather than too wide a range which, at worst, will sample extreme states of the world in which all alternatives fail”. Thus, in the context of informing robust decision making, it can be preferable to be slightly under- than slightly overconfident. To minimize the risk of producing overconfident projections we only use observational data with relatively uncontroversial and well-defined error structure.

\section*{Model setup}
\begin{figure}[!ht]
\begin{adjustwidth}{-1.75in}{0in}
\includegraphics{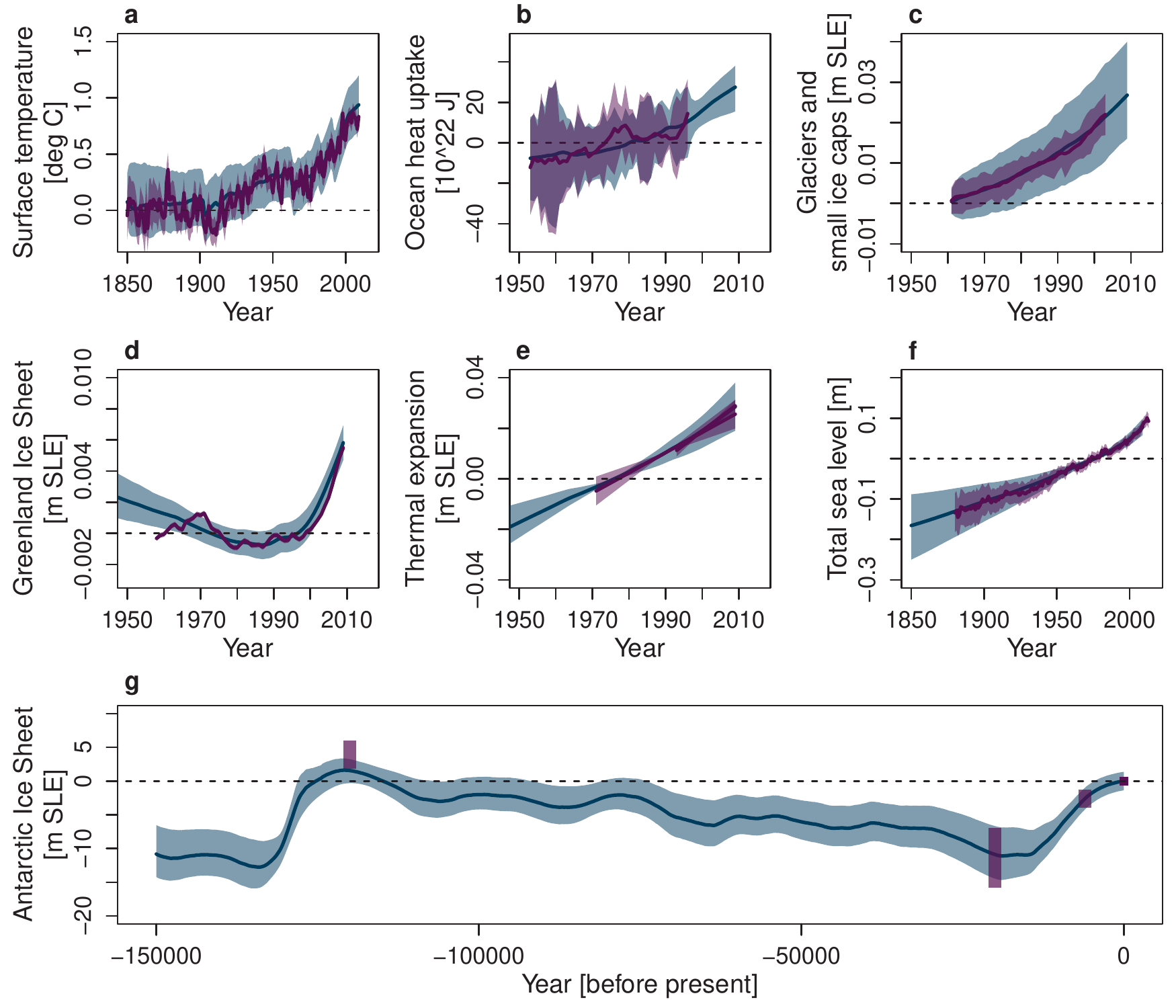}
\caption{{\bf Past observations (purple) and hindcasts (blue) global temperature, ocean heat content, sea-level contribution and global sea-level.}
Shadings represent the uncertainty (2\textsigma) in the observational data and the 5-95\% range in calibrated hindcasts.}
\label{fig1}
\end{adjustwidth}
\end{figure}

\paragraph{}
We use a relatively simple (39 free physical and statistical parameters), but a mechanistically motivated model framework to link transient sea-level rise to radiative concentration pathways applying sub-models for the global climate, Thermal Expansion (TE), and contributions of the Antarctic Ice Sheet (AIS), Greenland Ice Sheet (GIS) and Glaciers and Small Ice Caps (GSIC) (see Methods). This approach extends on the semi-empirical model setup recently reported by Mengel et al.~\cite{mengel2016}.
\paragraph{}
We use a Bayesian calibration method, wherein paleo-climatic data is assimilated with the AIS model separately from the calibration for the rest of the model, which assimilates only modern observations. Modern model simulations are then run at parameters drawn from the two resulting calibrated parameter sets (AIS and rest-of-model) and compared to global mean sea-level (GMSL) data~\cite{church2011} (see Methods). Only model realizations which agree with each GMSL data point to within 4\textsigma{} are admitted into the final ensemble for analysis. 4\textsigma{} was chosen so the spread in the model ensemble characterizes well the uncertainty in the GMSL data (Figure \ref{fig1}f).

\newpage

\paragraph{}
We choose, at this time, not to use paleo-reconstructions nor reanalyses, beyond incorporating a windowing approach into our calibration method for the Antarctic ice sheet parameters. This choice is motivated by the highly complex and uncertain error structure of these data sets. Failure to account for such complex error structure can result in considerable overconfidence, especially for low-probability events~\cite{ruckert2016a}.

\section*{Observational data and hindcasts}
\paragraph{}
Global temperature, ocean heat, and most sea-level contributions have typically been subject to upward, slightly accelerating trends since 1850 (Figure \ref{fig1})~\cite{church2013}. Only the sea-level contribution from AIS has been close to zero and might even have been slightly negative~\cite{church2013}. The reliability of the datasets decreases back in time due to the lower data availability and only the datasets for global mean surface temperature and global mean sea level go back to before 1950.
\paragraph{}
For the oceanic thermal expansion we use trends (together with the uncertainty estimates) as reported by the IPCC~\cite{church2013} for the calibration (Table S1). The time scale and uncertainties of the paleo-climatic AIS data are substantially different from the those of the observational data for other components of sea-level rise. For this reason, we calibrate the AIS model separately from the others, based on paleo-data as previously used by Shaffer~\cite{shaffer2014} and Ruckert et al.~\cite{ruckert2016b} (see Methods).
\paragraph{}
In general, the calibrated hindcasts (including both statistical and parameter uncertainty) correspond reasonably well to the reported uncertainty ranges surrounding the observational data. After calibration and \textit{post-calibration} our hindcasts of especially global temperature and global sea-level match the observations fairly well whereas the component models show some small deviations; GSIC uncertainty is slightly too large and the high AIS contribution during the Last Interglacial period is somewhat underestimated.

\section*{Projections}
\begin{figure}[!ht]
\includegraphics{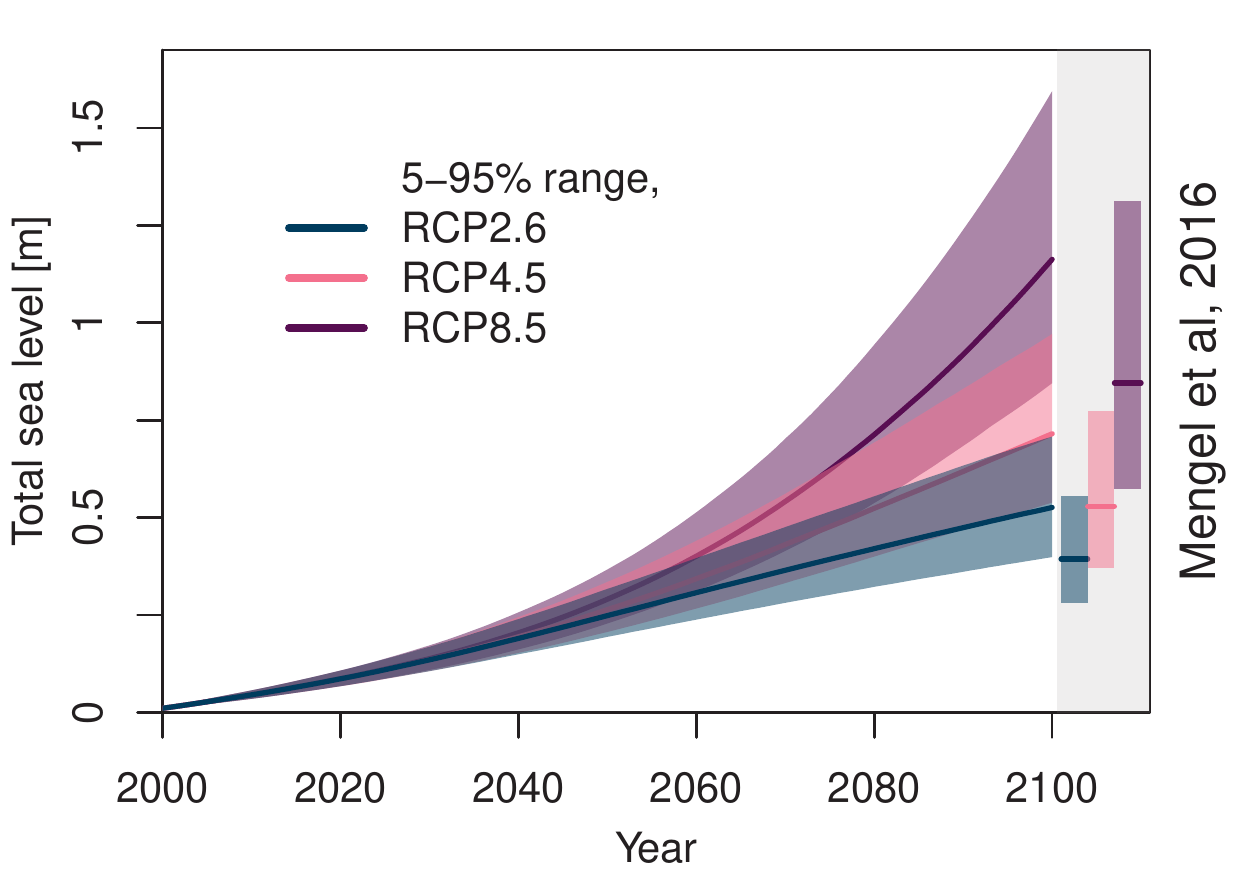}
\caption{{\bf Future probabilistic global sea-level projections for the 21st century under RCP2.6 (blue), RCP4.5 (pink) and RCP8.5 (purple) forcing scenarios~\cite{moss2010}, compared to the projections for 2100 by Mengel et al.~\cite{mengel2016} (vertical side bars)}}
\label{fig2}
\end{figure}

\begin{figure}[!ht]
\begin{adjustwidth}{-1in}{0in}
\includegraphics{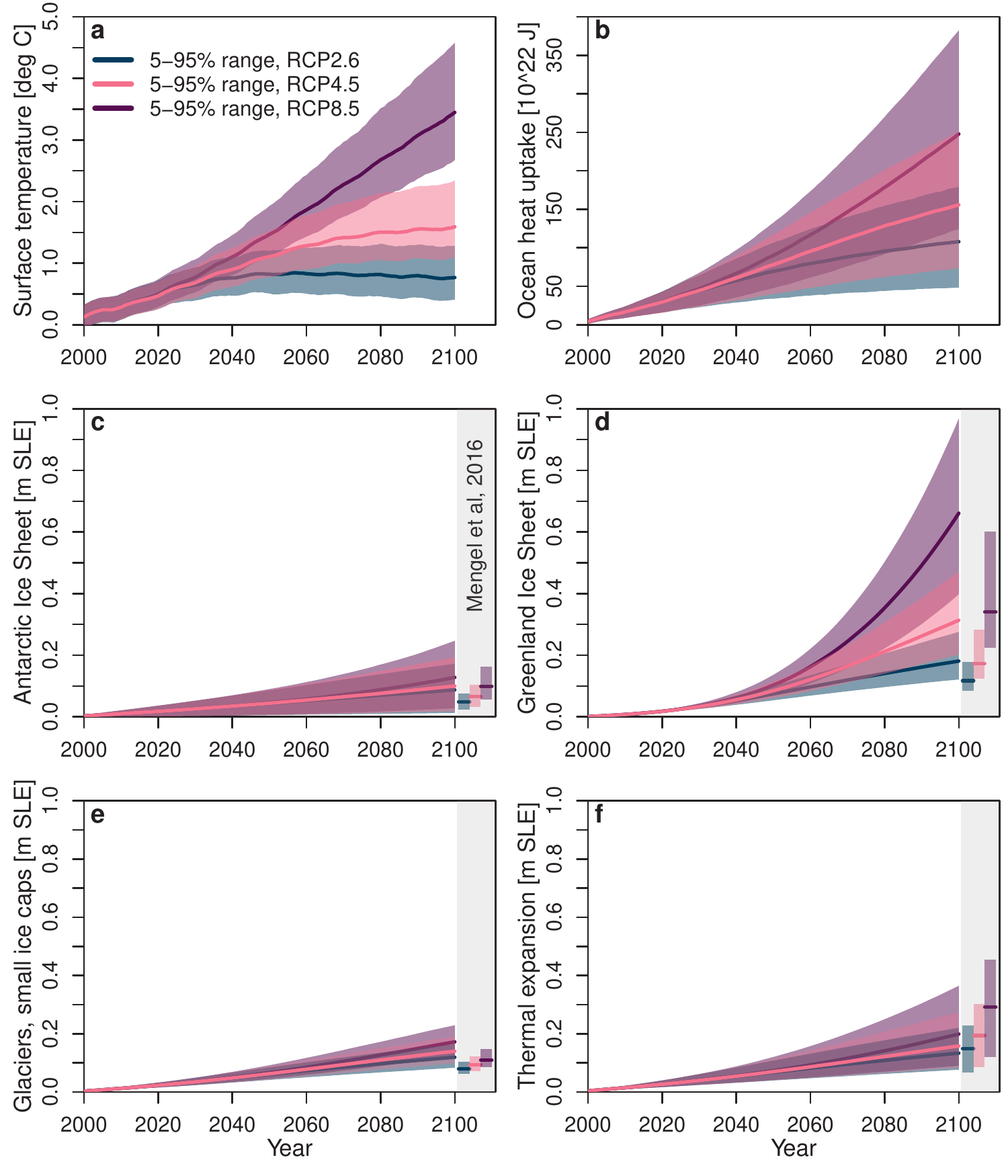}
\caption{{\bf Future probabilistic projections of global temperature, ocean heat content and sea-level contributions.}}
\label{fig3}
\end{adjustwidth}
\end{figure}

\paragraph{}
The probabilistic part of the projections yields 90\% credible ranges of 0.40-0.71 m (RCP2.6), 0.54-0.97 m (RCP4.5), and 0.85-1.59 m (RCP8.5) sea-level in 2100, relative to 1986-2005 mean sea level (Figure \ref{fig2}). This is slightly higher than projected by the recent and comparable study of Mengel et al.~\cite{mengel2016} 13 (that projects 0.28-0.56 m, 0.37-0.77 m, and 0.57-1.31 m for RCP2.6, RCP4.5, and RCP8.5, respectively) and can be explained by the relatively large contributions from the large ice bodies (Figure \ref{fig3} and Table S2). Our projected uncertainties in global sea-level are however quite similar to the results of Mengel et al.~\cite{mengel2016}.

\newpage

\paragraph{}
This similarity seems somewhat surprising since we deliberately aimed to be conservative with our prior parameter choices, but can be explained by our two-step calibration approach (see Methods). In the first step we calibrate the individual components of sea-level rise separately (similarly to Mengel et al.~\cite{mengel2016}), which indeed gives much wider uncertainty ranges in the projected sea-level rise and its components (not shown). Yet, those separate ranges are considerably reduced by the second combined calibration step based that also assimilates global sea-level data.

\section*{Deep uncertainties}
\begin{figure}[!ht]
\includegraphics{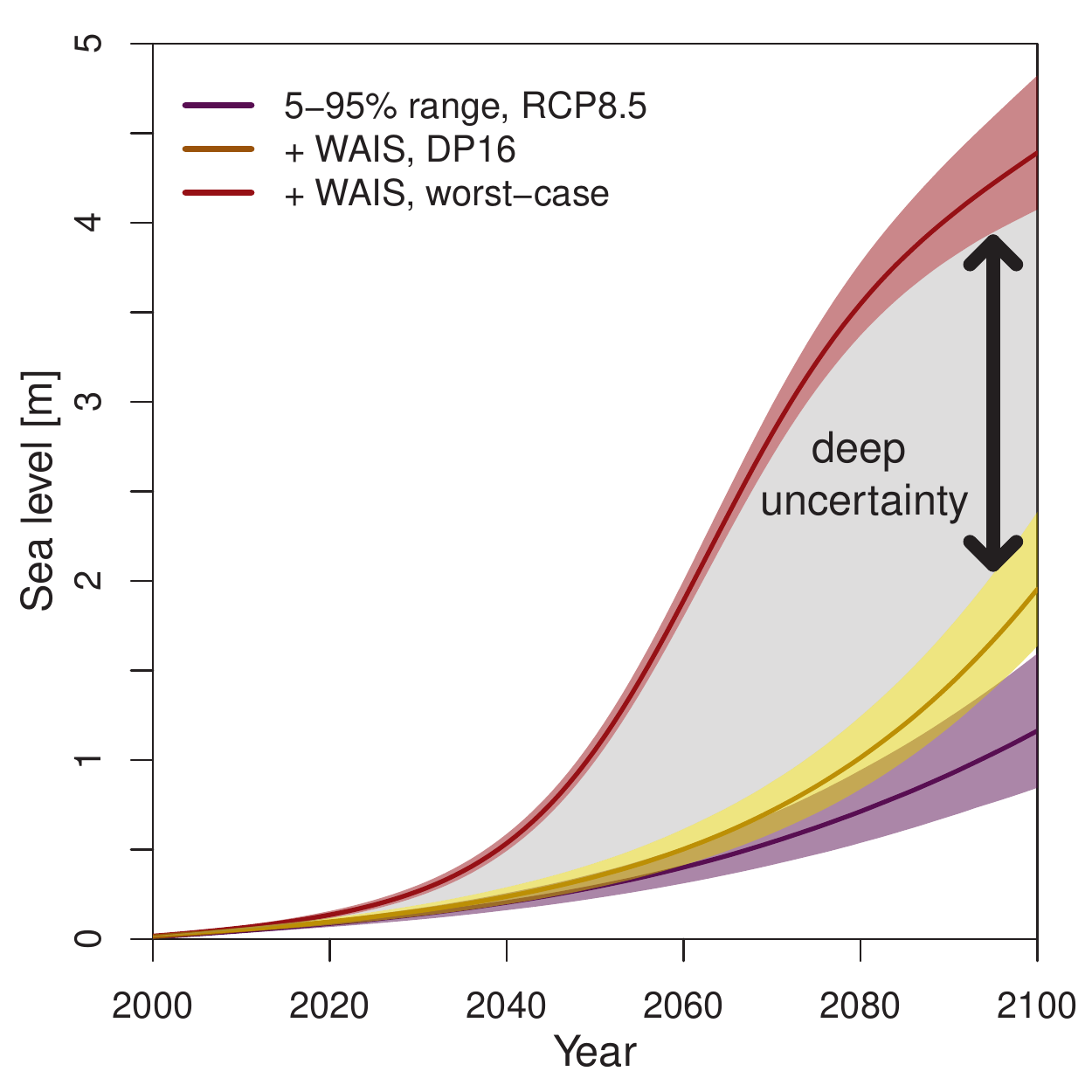}
\caption{{\bf Future sea-level projections including deeply uncertain contribution AIS response.}}
\label{fig4}
\end{figure}

\paragraph{}
Pollard et al.~\cite{pollard2015} suggests that a WAIS collapse might be possible on the order of decades. Yet, the timing of a rapid disintegration is deeply uncertain. DeConto and Pollard~\cite{deconto2016} present four widely divergent uncertainty ranges with, depending on the model choices, central estimates ranging from 64 to 114 cm for the sea-level contribution at 2100 following RCP8.5.
\paragraph{}
Our study contributes to communicating this deep uncertainty by characterizing the effect on plausible changes in global sea-level rise given the additional processes such as oceanic thermal expansion.  We provide three projections based on three WAIS-collapse scenarios, following RCP8.5; no collapse (0cm), a mid-range estimate (79cm, based on DeConto and Pollard~\cite{deconto2016}, see methods), and a high case (3.3m, full WAIS disintegration within a couple decades~\cite{pollard2015} (Figure \ref{fig4}). For 2100, this implies a factor two to four wider uncertainty range that should be accounted for to design robust strategies to cope with the deeply uncertain sea-level response to anthropogenic climate change. It is important to note that we do not intend to assign an implicit probability distribution to these deeply uncertain projections. We simply want to characterize and communicate key aspects of the deeply uncertain WAIS contribution to sea-level rise.

\newpage

\section*{Conclusions and discussion}
\paragraph{}
We presented a set of sea-level projections designed to represent important deep uncertainties and to inform robust decision-making frameworks. Our simple model framework includes semi-empirical models of the climate and sea-level contributions from thermal expansion, the Antarctic ice sheet, the Greenland ice sheet, and glaciers and small ice caps. Its relative simplicity is chosen to result in a transparent model structure and to enable a date-model fusion. Our calibration is designed to avoid overconstraining the projections. We hence only utilize observational data accompanied with clear uncertainty estimates, and aim for relatively non-informative prior distributions. We communicate divergent expert assessments and large structural uncertainties as deep uncertainties surrounding the projections.
\paragraph{}
The deeply uncertain contribution of WAIS disintegration dominates the overall uncertainty surrounding the sea-level projections. We present examples of low and high sea-level rise scenarios that could be expanded by relying more heavily on expert elicitation~\cite{bamber2013b,oppenheimer2016} or by incorporating strong priors on the characterization of the West Antarctic deep uncertainties.

\section*{Methods}
\subsection{Semi-empirical model framework}
\paragraph{}
We combine previously published, semi-empirical models (Table S1). The global temperature and ocean heat content are simulated with the coupled zero dimensional climate and 1D ocean model DOECLIM~\cite{kriegler2005}. The global mean surface temperature anomaly ($T_{g}$) feeds into the four models of sea-level contribution from thermal expansion (e.g.~\cite{mengel2016}), glaciers and small ice caps (submodel of MAGICC)~\cite{wigley2005}, the Greenland ice sheet (SIMPLE)~\cite{bakker2016a}, and the Antarctic ice sheet (DAIS)~\cite{shaffer2014}.

The DAIS model also requires Antarctic ocean surface temperatures ($T_{ANTO}$) which we estimate from a simple linear relation with $T_{g}$ bounded below at the freezing point of salt water ($T_{f}$ = -1.4\textdegree C),

\begin{center}
$T_{ANTO} = T_{f} + \frac{a_{ANTO} \times T_{g} + b_{ANTO} - T_{f}} {1+ exp[(a_{ANTO} \times T_{g} + b_{ANTO} - T_{f}) \ a_{ANTO}]}$
\end{center}
where $a_{ANTO}$ is the sensitivity of the Antarctic ocean temperature to global mean surface temperature (unitless), and $b_{ANTO}$ is the Antarctic ocean temperature for $T_{g}$=0\textdegree C. $a_{ANTO}$ and $b_{ANTO}$ are both estimated as uncertain model parameters.
\paragraph{}
For the models with four or fewer physical parameters (TE-model, MAGICC-GSIC, SIMPLE, and ANTO) we calibrate all parameters. For DOECLIM we apply the same free (physical) parameters as Urban et al.~\cite{urban2010} (climate sensitivity (\textit{S}), the aerosol amplification factor (\textalpha), and the ocean vertical diffusivity (\textkappa)), and for DAIS the same as used by Shaffer~\cite{shaffer2014} and Ruckert et al.~\cite{ruckert2016b}.

\subsection{Model calibration}
\paragraph{}
The model calibration approach consists of two stages. In the first stage, the AIS model is calibrated using paleo-climatic data as in Ruckert et al\cite{ruckert2016b}, along with trends in the AIS mass balance from the IPCC AR5~\cite{church2013}. The rest of the model components are similarly calibrated using modern observations. The reason for the separate calibrations is the vastly different temporal scale and characterization of errors between the paleo-climatic versus the modern data. All of the calibration data are detailed in Table S1. The model calibration is done using a robust adaptive Markov chain Monte Carlo (MCMC) approach~\cite{vihola2012}. For both the paleo-climatic and the modern calibrations, Gelman and Rubin diagnostics are examined to assess convergence~\cite{gelman1992}.
\paragraph{}
All parameters are assigned wide, physically-motivated prior ranges (Table S3), intentionally taken at least as wide as ranges considered in previous studies~\cite{ruckert2016b,urban2010} or divergent estimates from the literature (Table S4). We rely on published ranges, if these ranges are derived from data other than we use for the full calibration. For example, climate sensitivity is one of the parameters of our climate model, but published uncertainty ranges rely often on the same past observational data. Using those uncertainty ranges as prior would double-count the information content in the data. If independent priors are not available, we formulate priors that are constrained by our mechanistic understanding, and pre-calibration~\cite{Edwards2011}. This approach is one potential source of deep uncertainty, especially in case of limited availability of data to update the prior distribution. We are not aware of uncontroversial prior distributions for a potential rapid ice-sheet contribution of the West-Antarctic ice-sheet and we therefore restrict ourselves to a deeply uncertain range.
\paragraph{}
In the paleo-climatic calibration, four parallel MCMC chains of 500,000 DAIS model realizations each are sampled. The first 120,000 iterations of each is removed for burn-in, yielding 1,520,000 posterior parameter estimates for analysis. For the modern calibration, four parallel MCMC chains of 1,000,000 iterations each of the coupled DOECLIM-thermal expansion-GSIC-GIS model (modern calibration) are simulated. The last 500,000 iterations from each chain are used for analysis as the calibrated "rest-of-model" parameter estimates, yielding 2,000,000 posterior parameter samples for analysis. 
\paragraph{}
50,000 sample parameter sets are drawn from the DAIS and rest-of-model calibrated parameter sets. The entire parameter combination at which the models were run is preserved in this sampling. What is lacking at this stage is the joint rest-of-model and DAIS parameter distribution. The post-calibration step estimates this link by running the entire BRICK sea-level rise module (DOECLIM-ANTO-thermal expansion-GSIC-GIS-AIS) at these sampled parameter values. The parameter combinations are restricted to only those which yielded model realizations for global mean sea-level (GMSL) which matched data~\cite{church2011} to within a four-sigma window around all GMSL data points. The four-sigma range was chosen so as not to overconstrain, but still restrict the ensemble to simulations with a realistic representation of GMSL. Out of the 50,000 posterior samples, 5,612 post-calibrated model simulations are found. These served as the parameter samples for projections of GMSL. Projections to 2100 of GMSL and its components (thermal expansion, GSIC, GIS, and AIS) are made using Representative Concentration Pathways 2.6, 4.5, and 8.5~\cite{moss2010}. Experiments conducted using alternative windowing approaches for the GMSL post-calibration show little (at most five centimeters) variation in the 5-95\% ranges of projected sea-level rise in 2100.

\paragraph{Online Content}
Methods, along with any additional Extended Data display items and Source Data, are available in the online version of the paper; references unique to these sections appear only in the online paper.

\paragraph{Acknowledgements} 
This work was partially supported by the National Science Foundation through the Network for Sustainable Climate Risk Management (SCRiM) under NSF cooperative agreement GEO-1240507 and the Penn State Center for Climate Risk Management. Any conclusions or recommendations expressed in this material are those of the authors and do not necessarily reflect the views of the funding agencies.

\paragraph{Author contributions}
A. Bakker and K. Keller designed the research. A. Bakker designed the initial figures and wrote the first draft. T. Wong and K. Ruckert took care of the major part of the coding. All contributed to the final text.

\paragraph{Code availability}
The sea-level rise model (consisting of the subcomponents of sea-level rise used here, and including the modified AIS model with fast dynamics), the model calibration codes, and all post-processing codes are freely available from Tony Wong (twong@psu.edu). (After the review process at a peer-reviewed journal, a link to a GitHub site hosting the final code versions will be provided here.)

\bibliography{main}

\bibliographystyle{unsrt}

\end{document}